\documentclass[conference]{IEEEtran}
\usepackage{graphicx}
\usepackage{caption}
\usepackage{subcaption}
\usepackage{epstopdf}
\usepackage{multirow}
\usepackage{diagbox}
\usepackage{amssymb}
\usepackage{amsmath}
\usepackage{listings}
\usepackage[ruled, linesnumbered]{algorithm2e}
\usepackage{algorithmic}
\usepackage{lineno}
\usepackage{rotating}
\usepackage{color}
\usepackage{setspace}
\usepackage{float}
\usepackage{subcaption}
\usepackage{url}
\usepackage{color}
\usepackage[numbers]{natbib}
\usepackage{verbatim}
\begin{document}
%
\title{Optimal Deployments of Defense Mechanisms for the Internet of Things}

\author{\IEEEauthorblockN{Mengmeng Ge\IEEEauthorrefmark{1}, Jin-Hee Cho\IEEEauthorrefmark{2}, Charles A. Kamhoua\IEEEauthorrefmark{3}, Dong Seong Kim\IEEEauthorrefmark{4}}
\IEEEauthorblockA{
\IEEEauthorrefmark{1}School of Information Technology, Deakin University, Geelong, VIC, Australia\\
\IEEEauthorrefmark{2}Department of Computer science, Virginia Tech, Falls Church, VA, USA\\
\IEEEauthorrefmark{3}U.S. Army Research Laboratory, Adelphi, MD, United States\\
\IEEEauthorrefmark{4}School of Information Technology and Electrical Engineering, University of Queensland, Brisbane, QLD, Australia\\
mengmeng.ge@deakin.edu.au, jicho@vt.edu, charles.a.kamhoua.civ@mail.mil, dan.kim@uq.edu.au}\\
}

\maketitle

\begin{abstract}
Internet of Things (IoT) devices can be exploited by the attackers as entry points to break into the IoT networks without early detection. Little work has taken hybrid approaches that combine different defense mechanisms in an optimal way to increase the security of the IoT against sophisticated attacks. In this work, we propose a novel approach to generate the strategic deployment of adaptive deception technology and the patch management solution for the IoT under a budget constraint. We use a graphical security model along with three evaluation metrics to measure the effectiveness and efficiency of the proposed defense mechanisms. We apply the multi-objective genetic algorithm (GA) to compute the {\em Pareto optimal} deployments of defense mechanisms to maximize the security and minimize the deployment cost. We present a case study to show the feasibility of the proposed approach and to provide the defenders with various ways to choose optimal deployments of defense mechanisms for the IoT. We compare the GA with the exhaustive search algorithm (ESA) in terms of the runtime complexity and performance accuracy in optimality. Our results show that the GA is much more efficient in computing a good spread of the deployments than the ESA, in proportion to the increase of the IoT devices.
\end{abstract}


\IEEEpeerreviewmaketitle

\section{Introduction}
Internet of Things (IoT) is a large-scale network consisting of numerous interacting heterogeneous entities, including machines and/or humans~\cite{Roman2013ComNet, Roman2011Com, Sicari2015ComNet}. IoT devices often face severe resource constraints (e.g., sensors, mobile devices) in terms of their energy, computational power, and/or limited bandwidth. Due to the severe resource constraints, conventional, strong security mechanisms, requiring high computational overhead, are not applicable. Sophisticated attackers may easily compromise the IoT devices to penetrate into a target network and launch more severe attacks later on. The sophisticated attackers, often called {\em advanced persistent threat} (APT), refer to the attackers that are capable of collecting intelligence about a target, bypass multiple layers of defense mechanisms, and exploit the target's vulnerability to compromise it. In this paper, we consider two defense mechanisms to increase the security of the IoT networks and investigate how to deploy these defense mechanisms in an optimal manner. Such defense mechanisms may include: (1) deception techniques to divert attackers from real assets; and (2) security patch management solutions to reduce attack surface. 

Cyberdeception is used to add an additional layer of defense into conventional security solutions such as intrusion detection systems (or IDSs), firewalls, and/or endpoint anti-virus software. It allows defenders to capture and analyze malicious behaviors by luring the attackers into a decoy system within a network and interacting with attackers. As normal users do not know the existence of the decoy system, defenders will only get alerts caused by the malicious intrusions. A honeypot is one of common deception techniques in order to create a fake asset to protect valuable assets by diverting attackers. However, the management complexity and/or scalability issues of the honeypots hinder more active usage of them by enterprises. Modern deception technology employs the basic honeypot technology with automation techniques, which allows distributed deployment and update of decoy systems to achieve adequate coverage and high cost-effectiveness. In this work, we assume to use the modern deception technology and aim to evaluate the different ways of deploying decoys for providing highly cost-effective defense solutions.

Security patches are used to fix software vulnerabilities to prevent a system from possible exploits with the aim of reducing attack surface. However, the effective patch solution has challenged IoT devices. First, many manufacturers sell the IoT devices with no mechanism in place for automated patch updates. They are only in favor of usability but often neglect security in the design phase~\cite{Kolias2017Com} in order to gain quick access to the IoT market. This leaves the consumers with unsupported, vulnerable devices after the purchase. Second, some IoT devices (e.g., medical devices) use commercial off-the-shelf (COTS) software for the operating system or runtime environment. Patches for the COTS may not be installed without validating the patches by the manufacturers. To patch the IoT devices, the enterprises need to have agreements with the manufacturers that clarify the obligations for creating and evaluating the patches for the devices and also testing the patches for the COTS running on the devices during the lifecycle support. We will use heterogeneous IoT devices by different manufacturers and assume that the enterprises make agreements with the manufacturers to implement security patch solutions for the IoT products.

In our prior work~\cite{Ge2017JNCA}, we proposed a framework for modeling and assessing the security of the IoT. The framework is used to construct a graphical security model and a security evaluator with various security metrics to automate the security analysis of the IoT. Graphical security models (e.g., attack graphs (AGs)~\cite{Sheyner2002SP}, attack trees (ATs)~\cite{Saini2008JCSC}) have been widely used in network security assessment. In particular, an AG can depict all possible sequences of attackers' actions to compromise the target, while an AT explores possible ways an attack goal is attained by combining different types of attacks. However, the complexity of computing a complete AG is increasing exponentially and the construction of the AT is not scalable as well. Hence, we adopt a scalable graphical security model called the {\em multi-layer hierarchical attack representation model} (HARM)~\cite{Hong2016JNCA, Ge2017JNCA} which combines AGs and ATs to solve the complexity and scalability issues introduced by single-layered security models. 

To compute the optimal deployments of defense mechanisms, we use multi-layered HARM to evaluate the effectiveness and efficiency of the proposed deployment of the defense mechanisms and apply the multi-objective genetic algorithm (MOGA) to maximize network security while minimizing deployment cost. To the best of our knowledge, this work is the first that evaluates the combinations of the deception technology and the security patch solution and identifies an optimal setting to deploy these defense mechanisms for the IoT environments. This work has the following {\bf key contributions}:
\begin{itemize}
\item We developed a hybrid defense mechanism by combining the deception technology and software patch solution for the IoT environments and evaluated the proposed approach based on the graphical security model and security metrics.
\item We identified an optimal setting to deploy the developed defense mechanisms for the IoT under resource constraints. In this work, given the resource budget constraints, we solved an multi-objective optimization (MOO) problem using a genetic algorithm (GA) and compared its performance against the performance of the exhaustive  search algorithm (ESA) as a baseline model.
\item The proposed defense mechanism is generic in that it provides various ways of deploying defense mechanisms for the resource constrained IoT environments. Our proposed scheme is highly customizable to optimize multiple objectives to meet given system requirements. 
\end{itemize}
The rest of the paper is organized as follows. Section~\ref{relatedwork} presents related work on security optimization solutions and deception technology used for an IoT environment. Section~\ref{approach} describes our system model, defense mechanisms, and attack model considered in this work. Section~\ref{sec:optimal_defense_deployment} addresses our problem formulation and optimization steps to solve a given MOO problem followed by a case study in Section~\ref{case}. Section~\ref{sim} demonstrates our experimental results and discusses their overall trends. Section~\ref{limit} discusses the applicability and limitations of our work and future work directions. Lastly, Section~\ref{con} concludes our paper.

\section{Related Work} \label{relatedwork}
\textbf{Security optimization for IoT:} Little work has addressed a problem on how to optimally select defense mechanisms for the IoT. Rullo~\emph{et al.}~\cite{Rullo2017PS} developed a resource allocation mechanism to ensure the security of the IoT based on an a Stackelberg game. Based on the decisions from the attack-defense game, this work derived the best security resource allocation plan to achieve multiple system goals in terms of minimizing maximal risk, maximal criticality, energy consumption, and allocation cost. 
Rullo~\emph{et al.}~\cite{Rullo2017ICDCS} also took another approach to develop an optimal security resource allocation for an IoT network with mobile nodes based on GAs. Both works~\cite{Rullo2017PS, Rullo2017ICDCS} assume that an attacker can proceed a next attack after it can compromise a security resource. However, it is not realistic as highly sophisticated, stealthy attackers can directly compromise other components of a system while not being detected. Further, these works focus on device-level security but do not concern system-level security. 


\textbf{Cyber deception in IoT:} La~\emph{et al.}~\cite{La2016IoTJ} introduced a game theoretic model to analyze the security of honeypot-enabled IoT networks. They assumed that the attacker may deceive the defender with suspicious or seemingly normal traffic and used the honeypot-enabled intrusion detection component which reroutes the suspicious traffic to the honeypots as a defense mechanism. The interaction between the attacker and defender was modeled based on a Bayesian game with incomplete information. Anirudh~\emph{et al.}~\cite{Anirudh2017ICCCSP} used honeypots for online servers to mitigate Distributed-Denial-of-Service (DDoS) attacks launched from the IoT devices. Pa~\emph{et al.}~\cite{Pa2015WOOT} developed an IoT honeypot to emulate the IoT devices and capture Telnet-based attacks and designed the IoT sandbox to analyze these attacks against the IoT devices running different CPU architectures. Dowling~\emph{et al.}~\cite{Dowling2017ISSC} created a honeypot that simulates a ZigBee gateway and used it to capture attacks for a further analysis. 

Based on our literature review, no prior work has evaluated the performance of the modern deception technology on the IoT and proposed the hybrid approach combining different defense mechanisms to enhance the security of the IoT via graphical security models. In this work, we propose a graphical security model to evaluate the effectiveness and efficiency of the defense mechanisms using our devised evaluation metrics and identify optimal deployment solutions based on an MOGA.

\section{Preliminaries} \label{approach}
In this section, we describe our system model along with defense mechanisms and attack model considered in this work.

\subsection{System Model} \label{system}
We consider an IoT network consisting of servers, client machines (e.g., computers), and IoT devices~\cite{Gubbi2013FutureGenCom, Roman2013ComNet}. We assume that traditional defense mechanisms are in place on the IoT network, including IDSs, firewalls, and anti-virus software on the servers and client machines. The IoT network has a central patch management system to patch critical security vulnerabilities in the servers and client machines. This work focuses on how to deploy the deception technology in the network and the patch management solution for the IoT devices to defend against sophisticated attacks, as described in Section~\ref{attack}. The assumptions are made based on the realistic deployment scenario of the IoT network in the smart hospital~\cite{Trap2017Medjack}. We detail the example network scenario in Section~\ref{examplenet}.

\subsection{Defense Mechanisms} \label{solution}
We use two defense mechanisms, including the deception technology and security patch management solution, and their strategic deployment for adaptive defense based on the combination of those two defense mechanisms for the targeted IoT. 

\subsubsection{Deception Technologies}
Once attackers are inside an IoT network, they start probing to collect information to identify potential valuable assets and then move laterally in the network to launch attacks based on the information they gathered during the reconnaissance~\cite{trapx2017retail}. To successfully lure attackers, the following issues should be discussed: (1) where the decoy systems should be deployed; (2) what types of decoys should be employed; and (3) what level of authenticity of the decoys should be applied. In this section, we discuss these issues and the associated purchase cost associated with each deployment method. 

Modern deception technologies integrate honeypot technology, visualization, and automation technologies. There are several emerging vendors (e.g., Illusive Networks, Attivo Networks, TrapX, Cymmetria, TopSpin~\cite{DeceptionVendors}) which implement the modern deception technologies as well as provide support for IoT networks in various domains (e.g., smart home, smart office, health care). 

There are two types of decoys/traps utilized throughout a network: {\em emulation-based} and {\em full operating system (OS)-based}. Both emulation-based and full OS-based decoys can be autonomously created to fit within the environment with no changes to the existing IT infrastructure. They provide various types of interactive capabilities. Emulated decoys allow defenders to create a variety of fake assets (e.g., IoT devices, endpoints, servers, routers) and to provide a large-scale coverage across the network. Full OS-based decoys allow replication of the actual production devices to increase the possibility of engaging the attacker and to reveal the attacker's intention. To increase the overall chances attackers access decoys, the combinations of the decoys should closely resemble the real usage of the network devices and guarantee the decoy diversity. In addition, the decoys are suggested to be deployed in every VLAN of the network. 

Deception technologies are implemented in different ways by different vendors~\cite{Attivo2016, Ovum2017Attivo, Ovum2017TopSpin, trapx2017product}. The implementation usually consists of an intelligence center and various types of decoys. The intelligence center is used to create, deploy, and update a distributed decoy system, provide automated attack analysis, vulnerability assessment, and forensic reporting, and integrate with other prevention systems (e.g., security incident and event management platform, firewalls) to block attacks. The decoy system includes the decoys deployed across the network. The intelligence center can be purchased as a platform including hardware appliances and/or software. The decoys can be purchased individually with an annual license fee based on the number of servers, client machines, and IoT devices to be protected.

\subsubsection{Patch Management}
The IoT network often consists of various types of IoT devices produced by different manufacturers. The enterprises have annual maintenance contracts with these manufacturers for repairing and upgrading services. However, there is no patch management solution for the IoT devices. Therefore, for each type of the IoT devices, the enterprises can make additional annual agreements with the manufacturers for the patch management solutions with a certain amount of investment. In real-world situations, more complex cases may happen. For instance, different IoT devices produced by the same manufacturer can get the patch support in one contract. But some manufacturers cannot provide patches for the IoT devices. However, this work proposes an approach that provides an optimal selection of deployments of defense mechanisms. We leave the other issues mentioned above to be addressed in our future work.

\subsection{Attacker Model} \label{attack}
We mainly concern an outside attacker in this work. The attacker aims to obtain private information from an IoT network and sell it for its economic gain. Considering the real-world scenarios where attackers target a smart hospital environment~\cite{Trap2017Medjack}, we consider the following {\bf attack behaviors}:
\begin{itemize}
\item An attacker lacks knowledge on whether a decoy system exists or not. The attacker's capability to detect the deception depends on the knowledge gap between the attacker and the real system state. A smart attacker is assumed to have a high capability to recognize an emulated decoy but may not be able to detect a full OS-based decoy easily.
\item After the attacker interacted with a decoy, the attacker's behavior is monitored. If the attacker realized that the device is a decoy, it terminates its activities with the decoy immediately and attempts to find a new target to get into the network.
\item An attacker is capable of identifying exploitable, unpatched vulnerabilities or unknown vulnerabilities and compromising the vulnerable servers and/or client machines~\cite{Borbor2017DBSec}. 
\item An attacker is capable of exploiting unpatched vulnerabilities through hidden malware (e.g., re-packing, polymorphism) and using them to compromise other devices (e.g., installation of the backdoors for persistent attacks)~\cite{Trap2017Medjack}.
\end{itemize}

\section{Optimal Defense Deployment} \label{sec:optimal_defense_deployment}
In this section, we discuss our problem formulation and the steps to optimize the deployment of defense mechanisms.

\subsection{Problem Formulation} \label{formulation}
We define a deployment vector as the problem input and three metrics to evaluate the optimality and efficiency (or cost) of the deployment and define the following optimization problem. 

We use the definitions of an IoT network and the graphical security model in our prior work~\cite{Ge2017JNCA} and describe decoy nodes. The IoT network can be defined as $\mathit{IoT}$ $=$ ($S$, $T$, $V$) where $S$ is a finite set of subnets, $T$ is a finite set of nodes and $V$ is a finite set of vulnerabilities. For each node $t$ $\in$ $T$, we specify one existing attribute and add three attributes: $t_{\mathit{type}}$ $\in$ \{\emph{Web server} \emph{(Red Hat)}, \emph{Endpoint} \emph{(Win10)}, \emph{MRI} \emph{(Win7)}, \emph{Smart TV} \emph{(Samsung)}, $\cdots$ \}) specifies the type of the node; $t_{\mathit{decoy}}$ $\in$ $\{\mathit{False}, {True}\}$ indicates the node is either real or decoy; $t_{\mathit{pr}}$ $\in$ $(0,1]$ refers to the probability that an attacker will interact with the node and use it as a stepping stone to compromise other nodes; $t_{\mathit{cost}}$ is the deployment cost of the decoy (i.e., $0$ if $t_{\mathit{decoy}}$ $\equiv$ $\mathit{False}$). 

For $t_{\mathit{pr}}$, if the node is real, the attacker will have the probability of $1$ to exploit the vulnerabilities and then use it to compromise other nodes; if the node is a decoy, the attacker may figure out the trap after the interaction with the node and then will terminate its activities with the node; or it is deceived by the decoy and be diverted to other decoy nodes afterwards. Therefore, the full OS-based decoy will have a higher probability to deceive attackers than the emulated decoy as the attacker is more likely to interact with the full OS-based decoy without detecting the deception. It is also assumed that the attacker cannot detect the decoy without interacting with the decoy; this leads to $t_{\mathit{pr}} > 0$.

\subsubsection{Deployment Vector}
We assume the network is divided into several VLANs (i.e., subnets). All decoy nodes deployed across the network have different types. Let $Y_d$ denote the set of node types for deception deployment and $Y_p$ denote the set of the IoT types for patch management. We define the deployment vector as follows.

{\bf Definition 1.} The deployment vector $\mathit{dv}$ is defined as an integer valued vector. The function $o$ : $Y_d$ $\rightarrow$ $\{0, 1, 2\}$ describes the integer value for each type of the decoy deployment in the network. The function $q$ : $Y_p$ $\rightarrow$ $\{0,1\}$ describes an integer value of the patch solution for each type of IoT nodes in the network. Let $\mathit{dv}_d$ denote the deployment vector for the deception technology and $\mathit{dv}_p$ denote the deployment vector for the patch management. We denote the deployment vectors by:
\begin{eqnarray}
\mathit{dv}_d & =& (o(\mathit{type_1}), o(\mathit{type_2}), \cdots, o(\mathit{type_{|Y_d|}})) \nonumber \\
\mathit{dv}_p & =& (q(\mathit{type_1}), q(\mathit{type_2}), \cdots, q(\mathit{type}_{|Y_p|})) \\
\mathit{dv} &= & \mathit{dv}_d \cup  \mathit{dv}_p \nonumber
\end{eqnarray}
We assume at least one server decoy should be deployed in the network to engage the attackers. Therefore, for the server decoy, the integer value indicates that a specific type of server decoys is not deployed ($0$), emulated ($1$) or full OS-based ($2$). We use emulated decoys for client machines and IoT devices. Hence, the integer value indicates that a specific type of decoys is either deployed ($1$) or not ($0$). For the patch management, the integer value shows that a specific type of the IoT devices can be either patched ($1$) or not patched ($0$).

\subsubsection{Metrics}
We assume that an attacker may have one or more targets in the network for achieving its goal (e.g., stealing data stored in servers). For each target, the attacker may be able to find multiple attack paths to reach the target via one or multiple entry points. An attack path specifies a sequence of nodes that the attacker can compromise to reach the target node. We consider a set of attack paths $AP$ for the attacker to reach all the targets from all possible entry points. Each attack path $\mathit{ap}$ $\in$ $\mathit{AP}$ is a sequence of nodes along the path. We divide $\mathit{AP}$ into two sets: $\mathit{AP}_r$ representing a set of attack paths with real nodes as targets and $\mathit{AP}_d$ indicating a set of attack paths with decoy nodes as targets. As the attacker will terminate its activities with the decoy upon detecting the deception, $\mathit{AP}_r$ only contains the real nodes while $\mathit{AP}_d$ contains both real nodes and decoy nodes.


We use the following three {\bf metrics} to evaluate the effectiveness and efficiency of the deployed decoy system:
\begin{itemize} 
\item {\em Decoy Nodes Fraction} ($\mathit{DNF}$): This metric refers to the average fraction of decoy nodes among all nodes along the attack path towards a decoy target, representing the average decoy percentage along the attack path. Let $\mathit{ap}_{\mathit{dt}}$ denote a set of decoy nodes along an attack path. $\mathit{DNF}$ is measured by:
\begin{equation} \label{eq_DNF}
\mathit{DNF}(IoT, dv) = \frac{\sum \limits_{ap \in \mathit{AP}_d}{\frac{|\mathit{ap}_{\mathit{dt}}|}{|\mathit{ap}|}}}{|\mathit{AP}_d|}
\end{equation}
where higher $\mathit{DNF}$ is more desirable.
\item {\em Node Interaction Probability} ($\mathit{NIP}$): This metric indicates the average probability that an attacker will interact with nodes along a given attack path and eventually reach a decoy target. This metric reflects the average probability that the attacker is diverted to the decoy target along the attack path. $\mathit{NIP}$ is estimated by:
\begin{equation} \label{eq_NIP}
\mathit{NIP}(IoT, dv) = \frac{\sum \limits_{ap \in \mathit{AP}_d}{\prod\limits_{t \in ap}{t_\mathit{pr}}}}{|\mathit{AP}_d|}
\end{equation}
where higher $\mathit{NIP}$ is regarded as better performance.
\item {\em Fraction of Residual Cost} ($\mathit{RCF}$): This metric represents how much cost remains after subtracting one deployment related cost from the total cost. Here the total cost indicates the cost incurred for deploying all potential defense mechanisms. To estimate this cost, we assume that the manufacturer charges the patch management solution for the same type of IoT devices at a fixed price. We also assume that all decoys are sufficiently different showing high diversity, not to be easily detected by attackers. In order to obtain $\mathit{RCF}$, we consider three cost related to defense mechanism deployments as follows:
\begin{enumerate}
\item {\em Intelligence Cost} ($IC$): This cost incurs when purchasing the intelligence center as a platform and is considered as a constant cost in this work.
\item {\em Total Patch Management Cost} ($\mathit{PMC}$): This cost includes the total patch management cost for IoT devices and is obtained by:
\begin{equation} \label{eq_PMC}
\begin{split}
\mathit{PMC}(IoT, dv) = \sum_{\substack{\mathit{type} \in Y_p\\
q(\mathit{type}) \equiv 1}}
\mathit{pmc}_{\mathit{type}}
\end{split}
\end{equation}
where $\mathit{pmc}_\mathit{type}$ denotes the patch solution cost for one type of the IoT devices.
\item {\em Decoy Deployment Cost} ($\mathit{DC}$): This cost considers the deployment cost of all decoys and is given by:
\begin{equation} \label{eq_DC}
\begin{split}
\mathit{DC}(IoT, dv) = \sum_{\substack{t_{\mathit{decoy}} \equiv \mathit{True}\\
t_\mathit{type} \in Y_d\\o(t_\mathit{type})>0}}
t_{\mathit{cost}}
\end{split}
\end{equation}
\end{enumerate}
Based on these three costs above (i.e., $IC$, $PMC$, and $DC$), $\mathit{RCF}$ is computed by:
\begin{equation} \label{eq_RCF}
\begin{split}
\mathit{RCF}(IoT, dv) = \frac{\mathit{TC} - (\mathit{IC} + \mathit{DC} + \mathit{PMC})}{\mathit{TC}}
\end{split}
\end{equation}
where $TC$ is calculated by summing the associated costs for all potential defense mechanisms. Higher $\mathit{RCF}$ indicates lower actual deployment cost, which is more desirable.
\end{itemize}

\subsubsection{Problem Statement}
The problem we aim to solve is an MOO problem where the above three metrics (i.e., $\mathit{DNF}(\mathit{IoT}, \mathit{dv})$, $\mathit{NIP}(\mathit{IoT}, \mathit{dv})$, $\mathit{RCF}(\mathit{IoT}, \mathit{dv})$) should be maximized. 


Given that each deployment of the defense mechanisms entails purchase and maintenance cost, the optimization problem is to compute a set of {\em Pareto optimal solutions} (or Pareto frontier)~\cite{Marler2004Survey} that provide a reasonable balance between the effectiveness and efficiency of the deployments of the defense mechanisms. Let $\mathit{DV}$ denote all possible deployments of the defense mechanisms for a given IoT network and $\mathit{PP}$ $=$ \{$\mathit{DNF}(\mathit{IoT}, \mathit{dv})$, $\mathit{NIP}(\mathit{IoT}, \mathit{dv})$, $\mathit{RCF}(\mathit{IoT}, \mathit{dv})$ $|$ $\mathit{dv}$ $\in$ $\mathit{DV}$\} denote all possible values of the evaluation metrics for the given $\mathit{DV}$. We define the Pareto optimal solutions (or Pareto frontier) in our optimization problem as follows.

\textbf{Definition 2.} The Pareto frontier $\mathit{F}$ for the three-objective optimization problem is \{($\mathit{dnf}^*$, $\mathit{nip}^*$, $\mathit{rcf}^*$) $\in$ $\mathit{F}$\} $\iff$ $\nexists$ ($\mathit{dnf}$, $\mathit{nip}$, $\mathit{rcf}$) $\in$ $\mathit{F}$ such that $\mathit{dnf}$ $\geq$ $\mathit{dnf}^*$ $\land$ $\mathit{nip}$ $\geq$ $\mathit{nip}^*$ $\land$ $\mathit{rcf}$ $\geq$ $\mathit{rcf}^*$, and $\mathit{dnf}$ $>$ $\mathit{dnf}^*$ $\lor$ $\mathit{nip}$ $>$ $\mathit{nip}^*$ $\lor$ $\mathit{rcf}$ $>$ $\mathit{rcf}^*$ where * indicates an optimal solution for the objectives and each ($\mathit{dnf}^*$, $\mathit{nip}^*$, $\mathit{rcf}^*$) is a strongly nondominated solution.

\subsection{Optimization Steps}

\begin{figure*}[htb]
\centering
\includegraphics[width=0.8\textwidth]{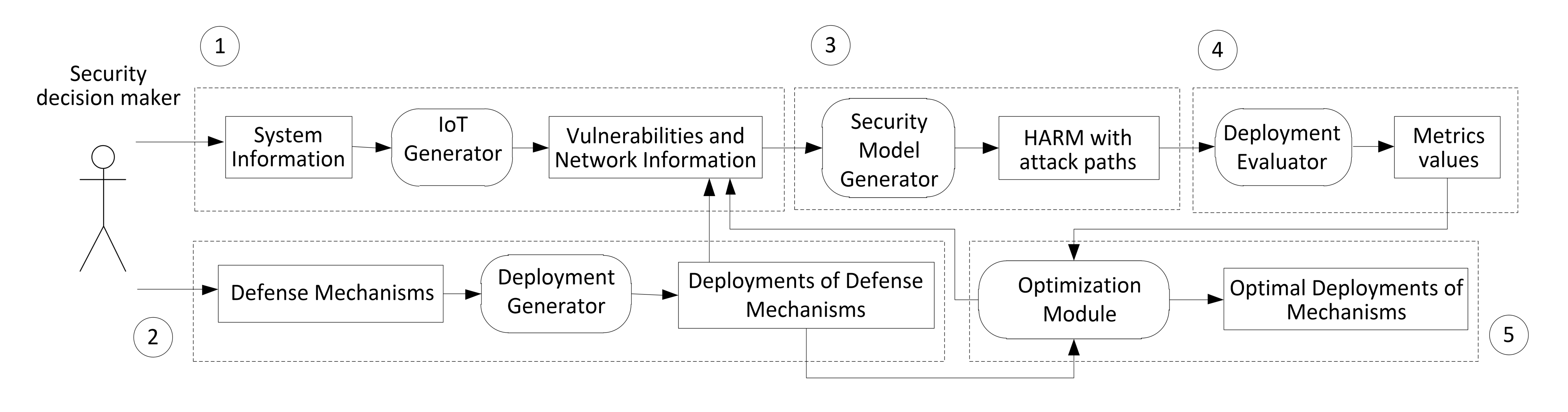}
\caption{Optimization steps in the proposed framework.}
\label{fig_flow}
\end{figure*}

We enhance our prior framework~\cite{Ge2017JNCA} by introducing the deployment generator, the deployment evaluator and the optimization module. Fig.~\ref{fig_flow} describes the enhanced, new framework with the following five phases: data input, deployment generation, security model generation, deployment evaluation, and deployment optimization. 

In {\em Phase 1}, the security decision maker provides the IoT Generator with the system information (i.e., network topology and node vulnerability information) to construct an IoT network. Given the network and all potential deployments of the defense mechanisms represented in the integer formats, the Deployment Generator randomly generates a set of different deployments in {\em Phase 2}. Each deployment of the defense mechanisms is fed into the network and also passed onto the Optimization Module. In {\em Phase 3}, the Security Model Generator takes the reconstructed network as input and automatically generates the HARM which captures all possible attack paths. We use the three-layered HARM~\cite{Hong2015TDSC} as our graphical security model. In the three-layered HARM, the upper layer captures the subnet reachability information, the middle layer represents the node connectivity information (i.e., nodes connected in the topological structure) and the lower layer denotes the vulnerability information of each node. In {\em Phase 4}, the Deployment Evaluator takes the HARM as input along with the evaluation metrics and computes the results which are then fed into the Optimization Module. In {\em Phase 5}, based on the initial set of deployments and the associated evaluation results, the Optimization Module applies the multi-objective genetic algorithm to compute the optimal deployments of the defense mechanisms for the IoT network based on the termination conditions (e.g., the maximum number of generations defined by the security decision maker).

We chose the GA to solve the given MOO problem due to the following reasons. First, GA provides a simple way to encode the candidate solutions. As we use the integer values to represent the deployment of defense mechanisms, binary encoding can be easily applied to convert the integer values into binary values for our defense mechanisms. Second, due to the rapidly growing IoT network, the scalability of an algorithm becomes a critical issue. However, GA requires little information to search effectively in a large search space which satisfies the requirement of an IoT network with a large number of nodes. To be specific, we choose one of the widely used GAs named {\em nondominated sorting GA II} (NSGA-II) in~\cite{Deb2002TEC} as the optimization algorithm. NSGA-II is defined as a fast sorting and elite MOGA and is able to find better spread of the solutions. 

\section{Case Study}
\label{case}
For our case study, we use an example scenario based on a smart hospital IoT environment. As a hospital system often keeps highly valuable information, attackers can have high economic gain by obtaining the private data. We will introduce the example network and explain each step to solve the given MOO problem based on our proposed approach. To deliver a more concrete idea, we are using this scenario to validate our proposed approach; however, our work is generic in nature and is applicable to general IoT environments.

\subsection{Example Network} \label{examplenet}

\begin{figure}[htb]
\centering
\includegraphics[width=0.35\textwidth]{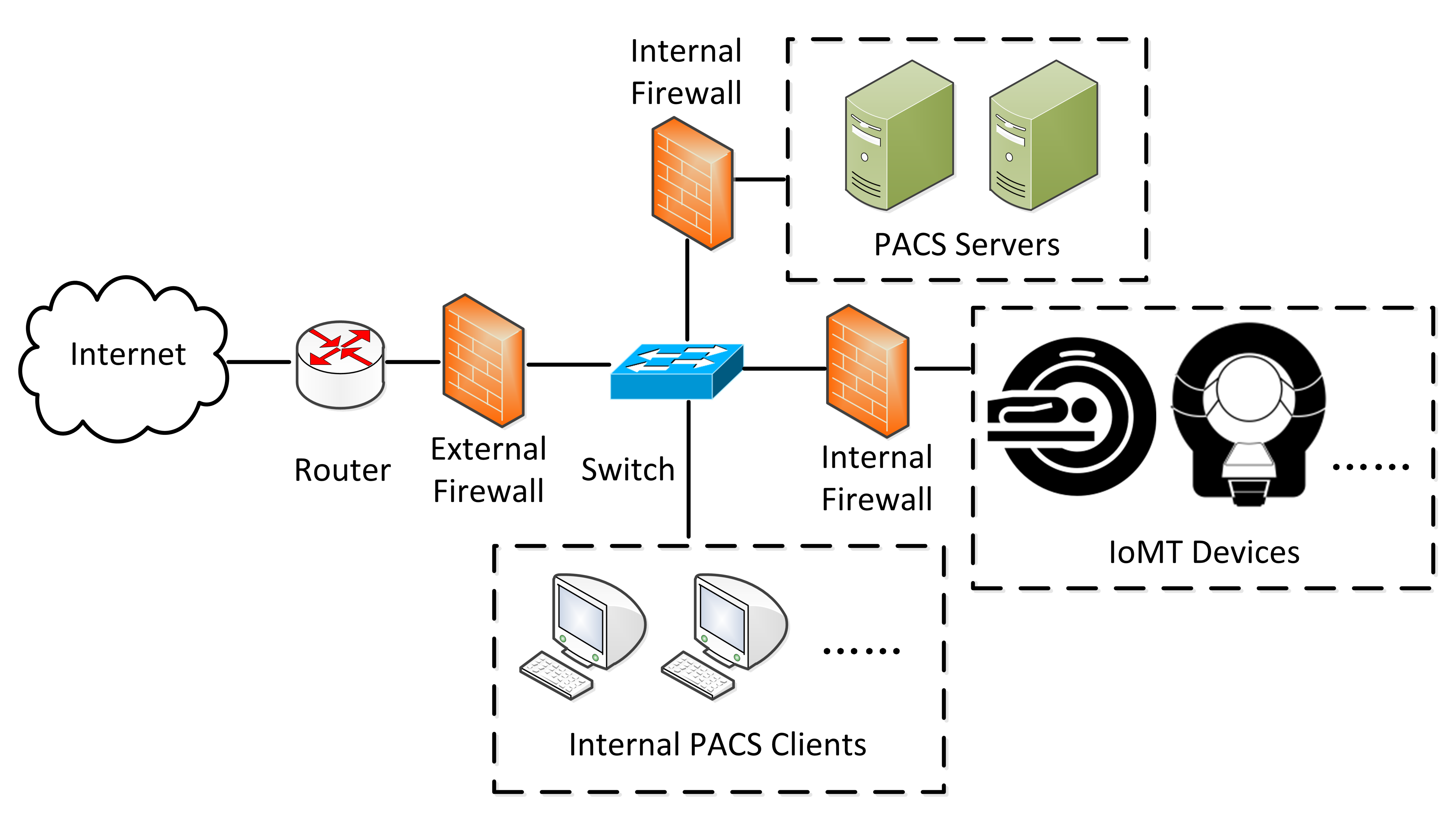}
\caption{An example PACS network.}
\label{fig_pacs}
\end{figure}

We consider the Picture Archive and Communication System (PACS) in a smart hospital IoT. The system consists of PACS servers for the storage of image information from multiple source machine types, PACS client machines to access the images and Internet of Medical Things (IoMT) using radiology techniques~\cite{Trap2017Medjack, Trap2017Medjack2} to send images to the servers. PACS uses the digital imaging and communications in medical (DICOM) standard as the communication protocol between the IoMT devices and the PACS servers. Fig.~\ref{fig_pacs} depicts the example PACS network.

The PACS network is divided into three VLANs by the external and internal firewalls. The PACS servers use active-active high availability cluster configuration to ensure reliable data storage and access. The redundant servers are identical in terms of both hardware and software. The PACS clients are computers with the DICOM viewers. The IoMT devices include Ultrasound machine, MRI machine, digital X-Ray machine, and CT scanner. As these IoMT devices are closed systems, additional security software cannot be easily integrated into the devices. Besides, patches for the COTS running on the IoMT devices need to be verified and tested by the manufacturers due to the safety issue.

We make assumptions for the operating systems (OSs) and applications running on the devices: each PACS server runs a Linux OS and is installed with the PACS software and database (e.g., MySQL); PACS client machines run two types of OSs, including Windows 8 and Windows 10; IoMT devices run Windows 7. All the chosen OSs and applications are commonly used in the PACS.

We use the attacker model described in Section~\ref{attack}. The attacker is able to wrap highly capable attack tools into obsolete malware (e.g., MS08-067), use it to exploit the vulnerabilities in the un-patched IoMT devices and become un-detected by the anti-virus software running on other devices~\cite{Trap2017Medjack, Trap2017Medjack2}. For the servers and client machines, the known but un-patched vulnerabilities can be collected from the National Vulnerability Database (NVD). We assume that the attacker is able to exploit the client machines and IoMT devices as entry points, then move laterally in different VLANs of the network, and eventually reach the servers for stealing private data.

The intelligence center is purchased as a platform and decoys are purchased individually with an annual license fee. The IoMT devices are purchased from different manufacturers. The hospital has the maintenance support from the manufacturers for the cleaning, repair and upgrade services but without the coverage of security updates. Thus, the hospital needs to make additional contracts with the manufacturers separately for the patch management support. We investigate the prices of the deception products~\cite{AttivoPrice, TrapXPrice} and service fees for the IoMT devices~\cite{Ultrasound, CTScan, MRI, XRay}. We consider that the additional patch management fee is 10\% of the average full service fee. The estimated prices for the defense mechanisms are shown in Table~\ref{tb_price}.
\vspace{2mm}
\begin{table}[htb] \scriptsize
\caption{Estimated prices for defense mechanisms.} \label{tb_price}
\centering
\begin{tabular}{|c|c|c|}
\hline
\multicolumn{2}{|c|}{Product} & Price (USD)\\
\hline
\multicolumn{3}{|l|}{$\bullet$ Deception Deployment}\\
\hline
\multicolumn{2}{|c|}{Intelligence Center} & 20,000\\
\hline
\multirow{2}{*}{PACS Server} & Full OS (Linux) & 1,500\\
\cline{2-3}
& Emulation (Linux) & 400\\
\hline
PACS Client & Windows 8 & 300\\
\cline{2-3}
(Emulation) & Windows 10 & 300\\
\hline
& Ultrasound & 200\\
\cline{2-3}
IoMT Device & MRI & 200\\
\cline{2-3}
(Emulation) & X-Ray & 200\\
\cline{2-3}
& CT Scanner & 200\\
\hline
\multicolumn{3}{|l|}{$\bullet$ Patch Solution}\\
\hline
& Ultrasound & 2,000\\
\cline{2-3}
IoMT & MRI & 6,000\\
\cline{2-3}
Device & X-Ray & 1,000\\
\cline{2-3}
& CT Scanner & 5,000\\
\hline
\end{tabular}
\end{table}

\subsection{Computation of the Optimal Deployments}

We consider a small-scale PACS network, including 2 PACS servers, 10 PACS client machines (5 running Windows 8 and 5 running Windows 10) and 4 different IoMT devices. The network is divided into three VLANs: servers in VLAN1, client machines in VLAN2 and IoMT devices in VLAN3. 

We assume each device has one known but un-patched vulnerability that can be exploited by the attacker to gain the root permission. Therefore, if the un-patched Windows 7 in the IoMT device is patched, there is no other exploitable vulnerability. We also assume each decoy has one vulnerability to lure the attacker. More vulnerabilities can be used based on the real configuration of the decoys. Besides, for an emulated decoy, the attacker interacts with the decoy with the probability of 0.5, can exploit the vulnerability and then use it to compromise other devices; for a full OS-based decoy, this probability is 0.9. We assume that the probability values are obtainable based on a measurement of the real configuration of the decoys (e.g., semi-quantitative and semi-qualitative).

Based on the concepts of GA~\cite{Marler2004Survey}, a population represents a group of potential deployments of the defense mechanisms; a chromosome corresponds to a deployment vector; a generation is one time of algorithm iteration; fitness values are determined by the evaluation metrics (i.e., fitness functions). We use the following algorithm parameters: population size $= 100$, maximum number of generations $= 100$, crossover rate $= 0.8$, and mutation rate $= 0.2$. 

We first generate the PACS network via the IoT Generator. The network is represented as $\mathit{IoT}_\mathit{PACS} = (S_\mathit{PACS}, T_\mathit{PACS}, V_\mathit{PACS})$ where 
\small{
\begin{eqnarray*}
S_\mathit{PACS} &=& \{s_{\mathit{VLAN1}}, s_{\mathit{VLAN2}}, s_{\mathit{VLAN3}}\}, \\
T_\mathit{PACS} &=& \{t_{\mathit{svr1}}, t_{\mathit{svr2}}, t_{\mathit{clt1}}, \cdots, t_{\mathit{clt10}}, t_{\mathit{iomt1}}, \cdots, t_{\mathit{iomt4}}\} \\ 
V_\mathit{PACS} &=& \{v_{\mathit{svr1}}, v_{\mathit{svr2}}, v_{\mathit{clt1}}, \cdots, v_{\mathit{clt10}}, v_{\mathit{iomt1}}, \cdots, v_{\mathit{iomt4}}\}.
\end{eqnarray*}
}
\normalsize
We show a list of attributes for $t_{\mathit{iomt1}}$ as an example: 
\small{ 
$$t_{\mathit{iomt1}_\mathit{type}} = Ultrasound (Win7); \;\;
t_{\mathit{iomt1}_\mathit{decoy}} = False;$$ 
$$t_{\mathit{iomt1}_\mathit{pr}} = 1.0;  \;\;
t_{\mathit{iomt1}_\mathit{cost}} = 0.$$
}
\normalsize
Given the network, we have the potential deployments of the defense mechanisms based on the types of devices: $Y_d$ = \{\emph{PACS server} \emph{(Linux)}, \emph{PACS client} \emph{(Win8)}, \emph{PACS client} \emph{(Win10)}, \emph{Ultrasound} \emph{(Win7)}, \emph{MRI} \emph{(Win7)}, \emph{XRay} \emph{(Win7)}, \emph{CT} \emph{(Win7)}\} and $Y_p$ = \{\emph{Ultrasound} \emph{(Win7)}, \emph{MRI} \emph{(Win7)}, \emph{XRay} \emph{(Win7)}, \emph{CT} \emph{(Win7)}\}.

Given the potential deployments, we randomly generate a population of 100 deployments $\mathit{DV}_\mathit{PACS}$. We show one example of the deployment vector, ${\mathit{dv}_1}$ = ${\mathit{dv}_1}_d \cup {\mathit{dv}_1}_p$ = $(o(\emph{PACS server (Linux)}), ...) \cup (q(\emph{Ultrasound (Win7)}), ...)$ = $(2, 1, 0, 1, 0, 0, 0) \cup (1, 0, 1, 1)$ = $(2, 1, 0, 1, 0, 0, 0, 1, 0, 1, 1)$.

\begin{figure}[htb]
\centering
\includegraphics[width=0.3\textwidth, height=0.325\textwidth]{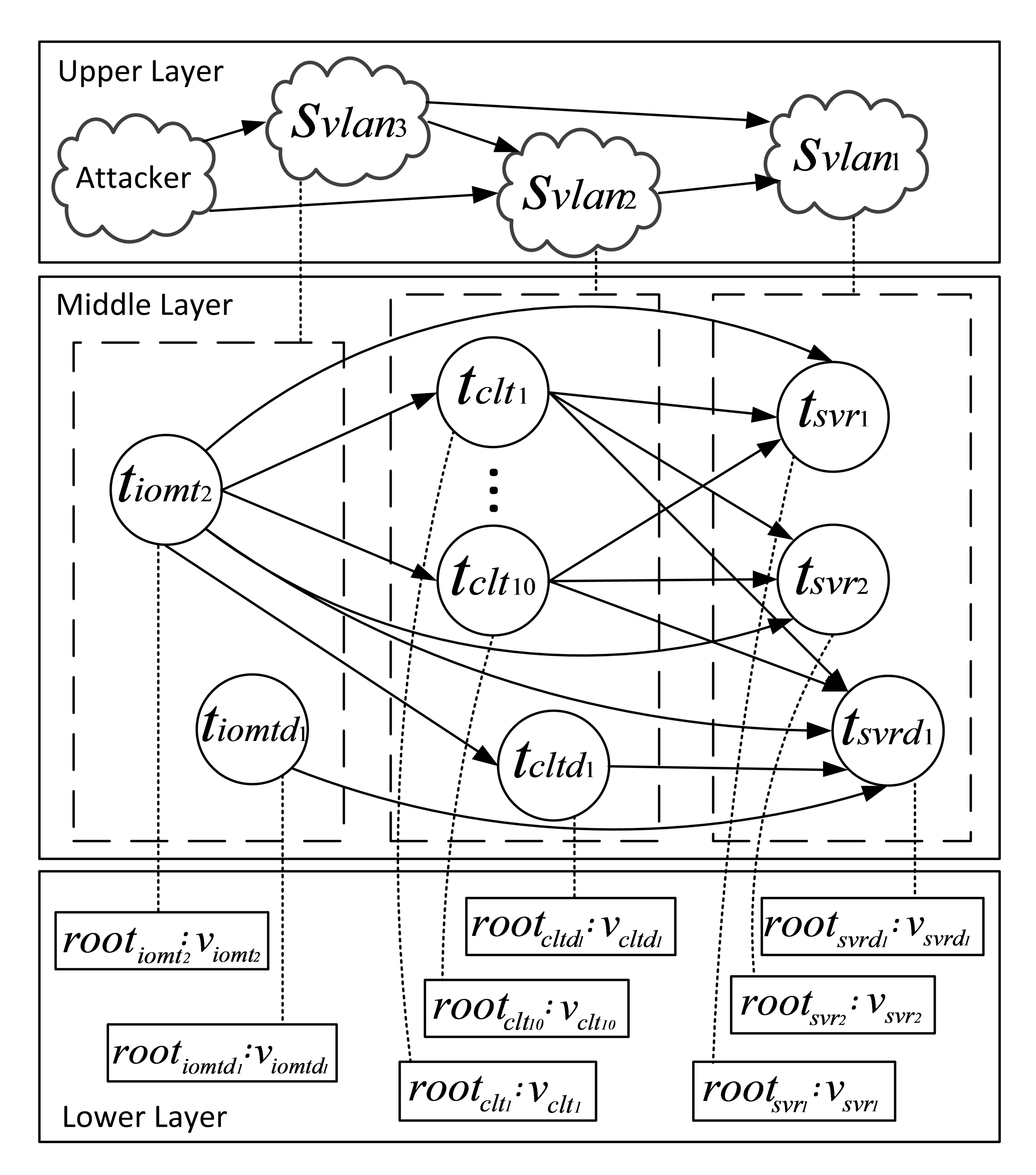}
\caption{HARM for the network with $\mathit{dv}_1$.}
\label{fig_harm}
\end{figure}

Afterwards, the PACS network is reconstructed with the defense mechanisms. For example, using $\mathit{dv}_1$, three decoy nodes are added into the network ($t_{\mathit{svrd1}}$, $t_{\mathit{cltd1}}$ and $t_{\mathit{iomtd1}}$) and three IoMT devices are patched ($t_{\mathit{iomt1}}$, $t_{\mathit{iomt3}}$ and $t_{\mathit{iomtd4}}$). The reconstructed network is fed into the Security Model Generator to construct the HARM and to capture the potential attack paths. We show the HARM for the network with the deployment vector $\mathit{dv}_1$ in Fig.~\ref{fig_harm}. 

From Fig.~\ref{fig_harm}, the attacker is able to take the client machines in VLAN2 and IoMT devices in VLAN3 as entry points and then move laterally in the network to eventually reach the servers. As decoys are also deployed in the network, the attacker may be lured into the decoys. Once the attacker interacts with the decoy, it either detects the decoy and terminates its attack behavior, or is diverted to another decoy. Then the HARM with the attack path information is taken as input into the Security Evaluator to evaluate the deployment of the defense mechanisms. We omit the calculation steps and results of the evaluation metrics for the network with $\mathit{dv}_1$ due to the page constraint.

\begin{figure*}[t!]
\centering
\begin{minipage}{0.33\textwidth}
  \centering
 \includegraphics[width=0.9\textwidth, height=0.6\textwidth]{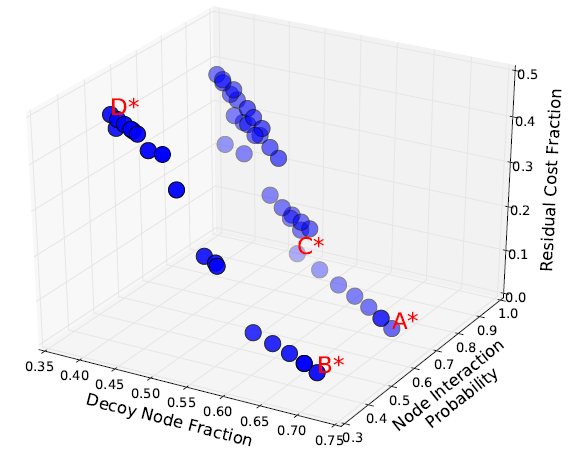}
\caption{Final deployments.}
\label{fig_case}
\end{minipage}%
\begin{minipage}{0.33\textwidth}
  \centering
\includegraphics[width=0.8\textwidth, height=0.6\textwidth]{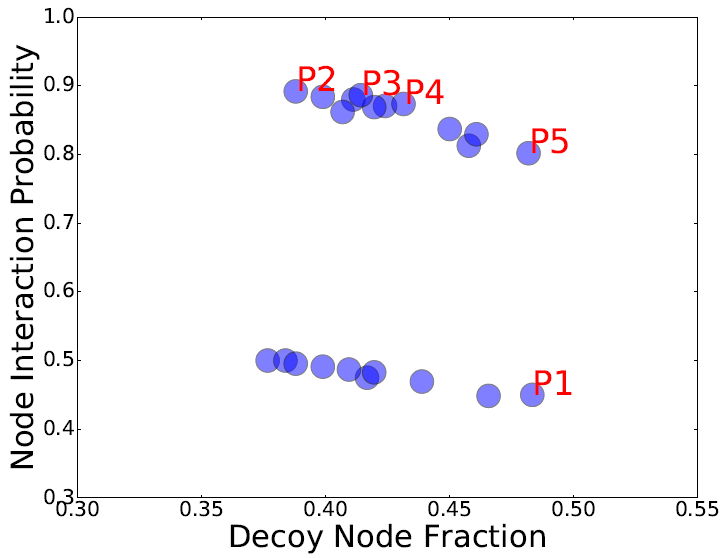}
\caption{Deployments with budget constraint.}
\label{fig_best}
\end{minipage}%
\begin{minipage}{.33\textwidth}
  \centering
\includegraphics[width=0.9\textwidth, height=0.6\textwidth]{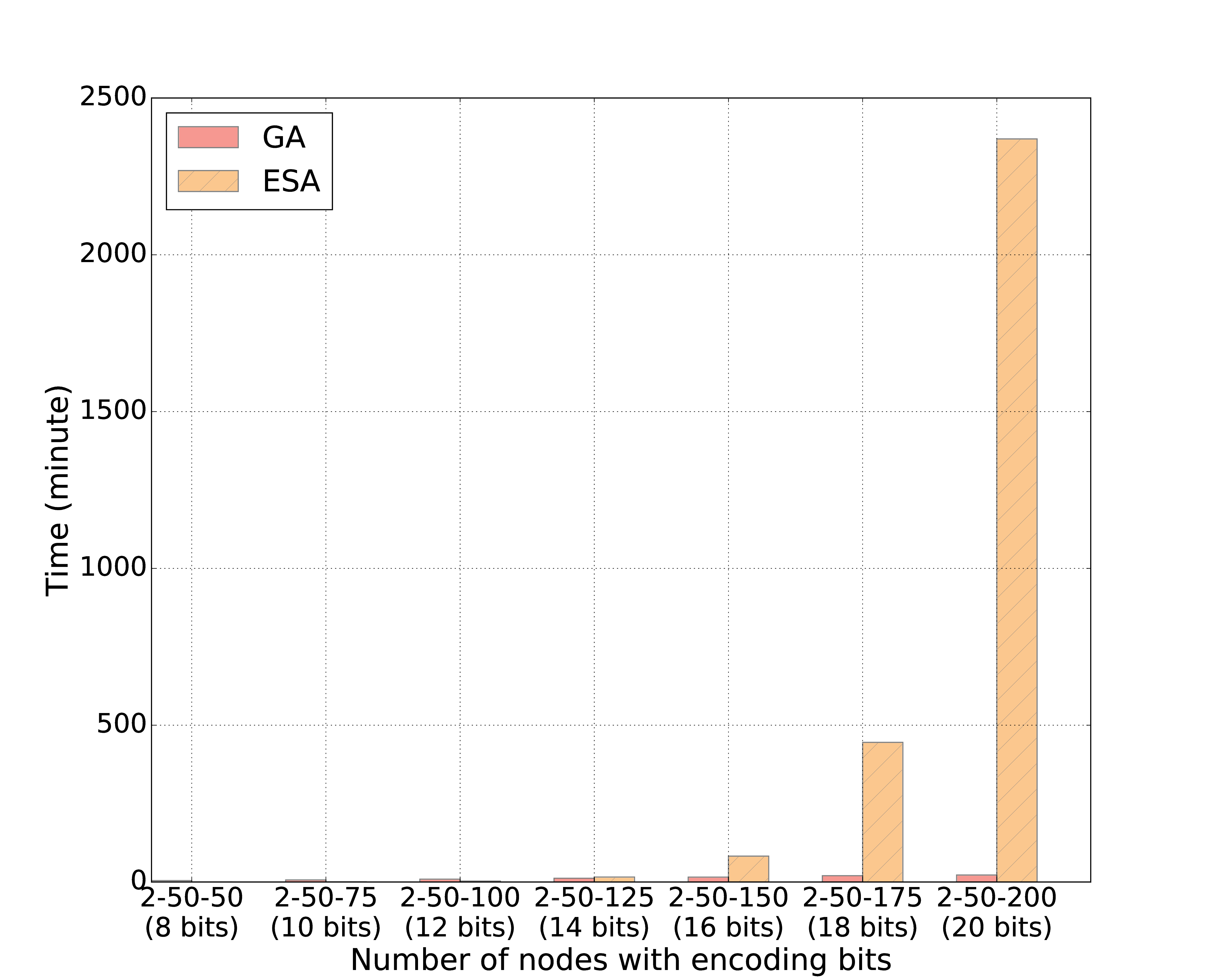}
\caption{Complexity: GA vs. ESA.}
\label{fig_time}
\end{minipage}
\end{figure*}

The Optimization Module takes the initial population of deployments ($\mathit{DV}_\mathit{PACS}$) and the corresponding fitness values ($\mathit{F}_\mathit{PACS}$) as inputs and then computes the optimal deployments via the MOGA. We show the final population of the fitness values in Fig.~\ref{fig_case} which forms the Pareto frontier ($\mathit{F}_\mathit{pacs}$). 

There are four labeled points in Fig.~\ref{fig_case} which represent the deployments with one maximum fitness value respectively. We show the deployment vector for each point as follows: ${\mathit{dv}_{A^*}}$ = (2, 1, 1, 1, 1, 1, 1, 1, 1, 1, 1), ${\mathit{dv}_{B^*}}$ = (1, 1, 1, 1, 1, 1, 1, 1, 1, 1, 1), ${\mathit{dv}_{C^*}}$ = (2, 0, 0, 0, 0, 0, 0, 1, 1, 1, 1) and ${\mathit{dv}_{D^*}}$ = (1, 0, 0, 0, 0, 0, 0, 0, 0, 0, 0).

Points $A^*$ and $B^*$ have the maximum decoy node fraction. In these two deployments, all decoys are deployed and all IoMT devices are patched. The only difference is that point $A^*$ represents the deployment with the full OS-based server decoy and point $B^*$ represents the one with the emulated server decoy. Point $C^*$ has the maximum node interaction probability and represents the deployment with the full OS-based server decoy deployed and all the IoMT devices patched. Point $E^*$ has the maximum residual cost fraction and represents the deployment with one emulated server decoy deployed. These deployments will be used to test the algorithm accuracy in Section~\ref{sim}.

\subsection{Analysis and Comparison of the Defense Mechanisms}
\label{subsec:opt-analysis}

In this section, we analyze the optimal deployments of the defense mechanisms and compare the optimal deployments with the deployments of the individual defense mechanisms. We introduce several metrics which are used in the following analysis: the percentage of decoys among all real devices ($\mathit{PD}$), the percentage of patched devices among all real devices ($\mathit{PPD}$), the number of attack paths towards the real targets ($\mathit{NAPRT}$ calculated by $|\mathit{AP}_r|$), the number of attack paths towards the decoy targets ($\mathit{NAPDT}$ calculated by $|\mathit{AP}_d|$) and the deployment cost of the defense mechanisms ($\mathit{DCDM}$). 

\subsubsection{Analysis of the Optimal Deployments}

We assume the hospital has a budget for the security investment and uses the budget and total cost to calculate the minimum residual cost fraction. Here, we use a budget of $\$25,000.00$ (in USD) and calculate the minimum $\mathit{RCF}$ which is approximately 0.322. We show the deployments meeting the budget constraint in Fig.~\ref{fig_best} (i.e., the points that are above 0.322 in Fig.~\ref{fig_case}).

To balance the effect of the two evaluation metrics in Fig.~\ref{fig_best}, we use a common scalarization method~\cite{Cho17}, which is a weighted sum to consider both metrics, $\mathit{DNF}$ and $\mathit{NIP}$, where each metric is weighted with $\beta$ and $\gamma$, respectively. This leads to a single, weighted metric, $\beta \mathit{DNF} + \gamma \mathit{NIP}$ ($\beta + \gamma = 1$). We use the weight of $\beta$ ranged from 0 to 1 (with the increment of 0.1) and show the points with the maximum weighted metric in Fig.~\ref{fig_best}. We summarize the corresponding deployment vector and the weight values of $\beta$ for each point in the following: 
${\mathit{dv}_{P_1}} = (1, 1, 1, 0, 1, 1, 0, 1, 0, 1, 0)$ with $\beta = 1$; ${\mathit{dv}_{P_2}}$ = $(2, 0, 0, 1, 0, 0, 0, 0, 0, 0, 0)$ with $\beta$ = 0, 0.1; ${\mathit{dv}_{P_3}}$ = $(2, 0, 0, 0, 0, 0, 1, 1, 0, 1, 0)$ with $\beta$ = 0.2, 0.3, 04; ${\mathit{dv}_{P_4}}$ = $(2, 0, 0, 1, 0, 0, 1, 1, 0, 1, 0)$ with $\beta$ = 0.5; ${\mathit{dv}_{P_5}}$ = $(2, 1, 1, 1, 1, 1, 1, 0, 0, 1, 0)$ with $\beta$ = 0.6, 0.7, 0.8, 0.9.

Point $P_1$ represents the deployment with the maximum decoy node fraction. Other points represent the deployments with one or multiple maximum weight metric values. We calculate the decoy percentage $\mathit{PD}$ and patch percentage $\mathit{PPD}$ for each point to indicate the coverage of the defense mechanisms. We show the values of two evaluation metrics ($\mathit{DNF}$ and $\mathit{NIP}$) and the percentages of each point in Table~\ref{tb_optimal} in the order of increasing $\beta$.
\vspace{1mm}
\begin{table}[htb] \scriptsize
\caption{Comparisons among the optimal deployments.}
\label{tb_optimal}
\centering
\begin{tabular}{|c|c|c|c|c|}
\hline
Point &$\mathit{DNF}$ & $\mathit{NIP}$ & $\mathit{PD}$ & $\mathit{PPD}$\\
\hline
$P_2$ & 0.388 & 0.892 & 12.5\% & 0.0\%\\
\hline
$P_3$ & 0.414 & 0.886 &12.5\% & 12.5\%\\
\hline
$P_4$ & 0.431 & 0.874 &18.8\% & 12.5\%\\
\hline
$P_5$ & 0.482 & 0.802 &43.8\% & 6.3\%\\
\hline
$P_1$ & 0.483 & 0.450 &31.3\% & 12.5\%\\
\hline
\end{tabular}
\end{table}

From point $P_2$ to point $P_4$ in Table~\ref{tb_optimal}, $\mathit{DNF}$, $\mathit{PD}$ and $\mathit{PPD}$ increase while $\mathit{NIP}$ decreases with the increasing value of $\beta$, which indicates the higher percentages of the decoys and the patched IoMT devices, and the higher decoy authenticity (deployment of the full OS-based server decoy) when the importance of $\mathit{DNF}$ increases. Point $P_5$ has the maximum $\mathit{PD}$, relatively high values of $\mathit{DNF}$ and $\mathit{NIP}$ and a low value of $\mathit{PPD}$, which demonstrates the maximum percentage of the decoys, high decoy authenticity but a low percentage of the patched IoMT devices. Point $P_1$ has the maximum $\mathit{DNF}$ and $\mathit{PPD}$, a low value of $\mathit{NIP}$ and a relatively high value of $\mathit{PD}$, which demonstrates the high percentages of the decoys and the patched IoMT devices but a low decoy authenticity (deployment of all emulated decoys).

Among the points $P_1$, $P_2$, $P_3$ and $P_4$, we can see that there is a balance between the decoy percentage and decoy authenticity. Point $P_1$ has a good percentage of the decoys (more decoys to trap the attackers) but low authenticity of the decoys (lower probability to interact with the attackers once they are diverted to the server decoy) while the other three points have the opposite effect. Point $P_5$ achieves a good percentage of the decoy and high authenticity of the decoys but has a low percentage of the patched IoMT devices. Besides, points $P_1$, $P_3$ and $P_4$ have a good patch percentage. In order to facilitate the decision making on the optimal deployment of the defense mechanisms by the defenders, we summarize the analysis results in the following way: (2) choose the deployment using ${\mathit{dv}_{P_5}}$ to achieve high decoy coverage and decoy authenticity; (2) choose the deployment using ${\mathit{dv}_{P_1}}$ to achieve high decoy coverage and patch coverage; and (3) choose the deployment using ${\mathit{dv}_{P_4}}$ to achieve high decoy authenticity and patch coverage.

\subsubsection{Comparison of the Defense Mechanisms}

We compare the three optimal deployments of the two defense mechanisms with the deployments of only deception or only patch solution in Table~\ref{tb_compare} using $\mathit{PD}$, $\mathit{PPD}$, $\mathit{NAPRT}$, $\mathit{NAPDT}$ and $\mathit{DCDM}$. Additionally, we patch all the IoMT devices by using only patch solution and deploy the full OS-based server decoy and all the other potential decoys by using only deception mechanism.
\vspace{1mm}
\begin{table}[htb] \scriptsize
\caption{Comparisons among the deployments.}
\label{tb_compare}
\centering
\begin{tabular}{|c|c|c|c|c|c|}
\hline
\diagbox[width=0.9in]{Defense}{Metric} & $\mathit{PD}$ & $\mathit{PPD}$ & $\mathit{NAPRT}$ & $\mathit{NAPDT}$ &$\mathit{DCDM}$\\
\hline
No defense & 0.0\% & 0.0\% &108 & 0 & 0\\
\hline
Only patch & 0.0\% & 25.0\% & 20 & 0 & 14000\\
\hline
Only deception & 43.8\% & 0.0\% & 108 & 68 & 22900\\
\hline
Both with ${\mathit{dv}_{P_1}}$ & 31.3\% & 12.5\% & 64 & 40 & 24400\\
\hline
Both with ${\mathit{dv}_{P_4}}$ & 18.8\% & 12.5\% & 64 & 34 & 24900\\
\hline
Both with ${\mathit{dv}_{P_5}}$ & 43.8\% & 6.3\% & 86 & 55 & 23900\\
\hline
\end{tabular}
\end{table}

From Table~\ref{tb_compare}, compared with no defense, different combinations of the defense mechanisms increase the security of the IoT network at different aspects. The deployment of patch solution has the highest patch percentage and lowest number of real attack paths. However, once the attacker breaks into the network, the real devices are the only targets and the behavior of sophisticated attackers may not be detected. The deployment of deception has the highest decoy percentage and highest number of fake attack paths but remains the highest number of real attack paths. The deployments of both patch and deception decrease the number of real attack paths significantly and introduce the fake attack paths to divert the attacks from the real assets. Therefore, the combinations of the two defense mechanisms are more effective to increase the security of the IoT networks with a reasonable cost.

Defenders could use our approach to compute the set of optimal deployments of the combined defense mechanisms and choose the most suitable deployment based on the budget constraint and analysis metrics.

\section{Simulations}
\label{sim}

We compare the GA with the exhaustive search algorithm (ESA) in terms of time efficiency and accuracy of the results. Here the accuracy refers to the ratio between the number of deployments with the maximum fitness value of one fitness function in the final population using the GA and that in Pareto frontier using the ESA. All simulations are performed using the computer equipped with a 3.4 GHz CPU under Linux Mint 18.1 Serena and Eclipse Neon.1 with Python 2.7.

We consider the IoT networks with the similar structure of the example network used in the case study along with the attacker model in Section~\ref{attack}. We use the network with a fixed number of servers and client machines and an increasing number of the IoT devices with different types to investigate the impact of the growing IoT devices on the proposed approach. Specifically, we use 2 servers, 50 client machines with two types of OSs. The number of IoT devices ranges from 50 to 200 with an increment of 25 in each simulation. Servers, client machines and different types of the IoT devices are deployed in different VLANs. We assume every 25 IoT devices have the same type and belong to the same manufacturer. Each manufacturer makes agreement with the enterprise to implement the patch solutions for the IoT devices at a different price (ranging from 1000 to 4000 with a difference of 500 for each agreement). Prices for the deception products are same as the prices shown in Table~\ref{tb_price}. We use the following algorithm parameters for the simulations: population size = 100, maximum number of generations = 100, crossover rate = 0.8 and mutation rate = 0.2. 

We show the time comparison of two algorithms in Fig.~\ref{fig_time}. We use the number of nodes with different types (in the form of server-client machine-IoT) along with the number of bits used for the binary encodings of the deployments as labels. Initially, when the scale of the network is small, the ESA runs faster than the GA. However, with the increasing number of the encoding bits, the time of the ESA increases exponentially while the time of the GA increases linearly.

We show the accuracy ratios of the GA compared with the ESA in Table~\ref{tb_accuracy}. When the number of the encoding bits ranges from 8 to 14, the accuracy ratio is 1.0 as the set of the deployments with one maximum fitness value calculated by the GA is equal to the set of deployments with one maximum value calculated by the ESA. The ratio decreases when the number of the encoding bits is equal to or larger than 16. We increase the population size and the maximum generation of GA and run the simulations with the same networks. For the network with 2 servers, 50 client machines and 150 IoT devices, we keep the current population size while increase the maximum generation to 150; for the network with 175 IoT devices, we increase the population size and the maximum generation to 150; for the network with 200 IoT devices, we increase the population size to 150 and the maximum generation to 200. We obtain the ratios of 1.0 for all these networks with a slightly higher runtime.  
\vspace{2mm}
\begin{table}[htb] \scriptsize
\caption{Accuracy ratios of GA.}
\label{tb_accuracy}
\centering
\begin{tabular}{|c|c|c|c|c|}
\hline
\multirow{2}{*}{Network} &Population & Maximum & Runtime & Accuracy\\ 
& size & generation & (minute) &ratio\\
\hline
2-50-50 (8) & 100 & 100 & 4 & 1.0\\
\hline
2-50-75 (10) & 100 & 100 & 7 & 1.0\\
\hline
2-50-100 (12) & 100 & 100 & 9 & 1.0\\
\hline
2-50-125 (14) & 100 & 100 & 12 & 1.0\\
\hline
\multirow{2}{*}{2-50-150 (16)} & 100 & 100 & 16 & 0.75\\
\cline{2-5}
& \textbf{100} & \textbf{150} & \textbf{22} & \textbf{1.0}\\
\hline
\multirow{2}{*}{2-50-175 (18)} & 100 & 100 & 20 & 0.75\\
\cline{2-5}
& \textbf{150} & \textbf{150} & \textbf{40} & \textbf{1.0}\\
\hline
\multirow{2}{*}{2-50-200 (20)} & 100 & 100 & 22 & 0.5\\
\cline{2-5}
& \textbf{150} & \textbf{200} & \textbf{67} & \textbf{1.0}\\
\hline
\end{tabular}
\end{table}

In conclusion, the GA is much more efficient than the ESA when the size of the IoT network increases and is able to obtain a good spread of the optimal deployments within a reasonable time. 

\section{Discussions}\label{limit}
The proposed approach can be applied to any IoT networks with the potential deployments of these two defense mechanisms. However, by designing new evaluation metrics and integrating them with the graphical security model, our approach can be used to investigate any combinations of the defense mechanisms. In our future work, we could use more case studies to analyze the effect of the defense mechanisms and apply game theoretic approach to model the interaction between the defender who tries to deploy optimal defenses and attacker who tries to strategically evade the defenses~\cite{Kamdem2017ICDCSW}. Besides, several limitations can be resolved to extend the scope of the work.
\begin{itemize}
\item \textbf{Deployment cost:} 
We will consider the long-term effect of the annual fees on the deployment costs of both the deception and security patch mechanisms.

\item \textbf{Deception:} The attacker may not choose the particular decoy IoT device among all IoT devices. We will investigate the effectiveness and efficiency of the deployments with the various numbers of decoys for each IoT type.

\item \textbf{Optimization algorithms:} We will validate whether optimal solutions are identified using GA via simulations and evaluate and compare other heuristic algorithms to find the most efficient algorithm for our optimization problem.

\item \textbf{Validation:} We introduced a case study to demonstrate the feasibility of our proposed approach and carried out simulations to evaluate the efficiency of the optimization algorithm. We will perform the experiment on a small-scale IoT network to verify the approach.  
\end{itemize}
\section{Conclusions} \label{con}
In this paper, we have provided a novel approach to compute the optimal deployments of defense mechanisms for IoT environments. We have introduced two defense mechanisms, including the modern deception technology and patch management solution for IoT devices. We have defined three metrics to measure the effectiveness and efficiency of the deployments of defense mechanisms and formalized a multi-objective optimization (MOO) problem to maximize three devised metrics as system goals. We have leveraged the genetic algorithm to solve the MOO problem and integrated it with the graphical security model along with the deployment evaluator in the framework to carry out the computation of optimal deployments of defense mechanisms. We have validated the performance of the proposed approach based on a case study with a smart hospital IoT scenario. Through the extensive simulation experiments, we have validated the performance of the proposed framework and deployment method by conducting a comparative analysis between the genetic algorithm and the exhaustive search algorithm in terms of runtime complexity and optimality (or accuracy). Our simulation results have shown that the GA generates high quality solutions with low complexity, by showing a good spread of the deployments of defense mechanisms. 

\linespread{0.9}
\bibliographystyle{IEEETranSN}
{\small
\bibliography{IoTsecurity}
}

\end{document}